%% file: main.tex
\documentclass[a4paper,11pt]{article}
\usepackage{pos}
\usepackage{lineno}
\title{Multimessenger NuEM Alerts with AMON}
 \ShortTitle{Multimessenger NuEM Alerts with AMON}

\author*[a]{Hugo Ayala}

\affiliation[a]{Pennsylvania State University,\\
  State College, US}

\forColl{AMON Group, the HAWC, the IceCube and the Antares} 

\emailAdd{hgayala@psu.edu}

\abstract{The Astrophysical Multimessenger Observatory Network (AMON), has developed a real-time multi-messenger alert system. The system performs coincidence analyses of datasets from gamma-ray and neutrino detectors, making the Neutrino-Electromagnetic (NuEM) alert channel. For these analyses, AMON takes advantage of sub-threshold events, i.e., events that by themselves are not significant in the individual detectors. The main purpose of this channel is to search for gamma-ray counterparts of neutrino events. We will describe the different analyses that make-up this channel and present a selection of recent results.}

\FullConference{37$^{\rm{th}}$ International Cosmic Ray Conference (ICRC 2021)\\
		July 12th -- 23rd, 2021\\
		Online -- Berlin, Germany}

\begin{document}
\maketitle

\section{Multi-messenger Astrophysics}

The study of the Universe has been benefited by the improvement of our technology and analysis techniques. Recently, this has allowed us to combine the information available from different observations that focus on individual components of the physical phenomena. The most important results in the past lustrum are the detection of gravitational waves and electromagnetic radiation from the merger of a binary neutron star~\citep{bns}, together with the detection of high-energy neutrinos and gamma-rays from the blazar TXS 0506-056~\citep{icneutrino}.

The information given by each individual messenger --- cosmic rays, electromagnetic radiation, neutrinos, and gravitational waves --- help us get a general picture about the acceleration mechanisms and identification of sources that produce cosmic-rays, gamma-ray bursts, the nature of dark matter and other unsolved questions in astrophysics \citep[See for example][]{mmareview, mmareview2}.

In these contribution we present an overview of the Neutrino-Electromagnetic (NuEM) channel of the Astrophysical Multimessenger Observatory Network. This effort is based on the new paradigm of multi-messenger astrophysics, which aims at coordination between different observatories and combining independent datasets into a coincidence analysis.

\section{AMON}
The Astrophysical Multimessenger Observatory Network (AMON) is a program developed at the Pennsylvania State University. Its primary objective is to perform real-time coincidence searches of sub-threshold events of different observatories. Any statistically significant coincidence is then reported to the astrophysical community through Gamma-ray Coordinates Network (GCN, also known as the Transient Astronomy Network).
AMON also stores events into its database to perform archival coincidence searches. It broadcasts individual events to GCN if the observatories providing the events consider them to be of interest to the astrophysical community. And finally, it aims to be a framework that facilitates the interaction between observatories trying to combine their datasets~\citep{amon_2020}. A list of participants can be found in the AMON webpage\footnote{\url{https://www.amon.psu.edu/amon-participants/}}.

The AMON alert system started sending alerts in 2016. Currently, AMON provides the following public alerts: 
\begin{itemize}
\item IceCube Gold, Bronze and Cascade alerts,
\item HAWC Burst-like alerts,
\item NuEM channel alerts.
\end{itemize}
The notices of these alerts can be found in the GCN webpage\footnote{\url{https://gcn.gsfc.nasa.gov/amon.html}}. The webpage also includes the now decomissioned IceCube HESE and EHE alerts.

\section{The NuEM Channel}

The AMON NuEM channel focuses on neutrino and high-energy photon coincidences by using sub-threshold data\footnote{The numerical definition of a sub-threshold event is set by each individual observatory.} from different observatories. The main objective of this channel is to search for the sources of high-energy neutrinos with the help of high-energy gamma rays. 

This search can be performed since there are physical processes where neutrinos and gamma rays are produced together. It is known that accelerated cosmic rays interact with radiation fields or interstellar matter surrounding the region around astrophysical sources. These interactions produce secondary charged and neutral pions. Charged pions mainly decay via $\pi^{+} \rightarrow \mu^{+} + \nu_{\mu}$, followed by the decay of the muon as $\mu^{+} \rightarrow e^{+} + \nu_e + \bar{\nu}_{\mu}$ and a similar process with the charge conjugate. Neutral pions decay into two gamma-ray photons, $\pi^{0} \rightarrow \gamma + \gamma$. A photohadronic interaction, i.e. a collision between a cosmic ray and a photon, will produce charged and neutral pions with probabilities of one-third and two-thirds. If the interaction is just hadronic, then the probability of producing charged and neutral pions is one-third for each type of pion~\citep{neutrinoProd}. 

The NuEM channel receives data from the IceCube \citep{icecube}, ANTARES \citep{antares}, HAWC \citep{hawc}, and \textit{Fermi}-LAT \citep{fermi} observatories.
We have performed archival coincidences analysis between: \textit{Fermi}-ANTARES \citep{antares_fermi}, HAWC-IceCube \citep{hawc_icecube}, \textit{Fermi}-IceCube \citep{icecube_fermi} and HAWC-ANTARES, which preliminary results are shown in this contribution. 

The general algorithm used in AMON to find and rank coincidences is as follows:
\begin{enumerate}
    \item Set a criteria to select events which conform a coincidence. The criteria can be:
    \begin{itemize}
        \item A time window where events can arrive. 
        \item A maximum angular distance between events.
    \end{itemize}
    \item Calculate a ranking statistic for the coincidence. This usually includes:
    \begin{itemize}
        \item A likelihood calculation, $\lambda(\boldsymbol{x})$, quantifying the overlap between events. Maximizing the likelihood, $\lambda_{\rm max}$, gives an estimate of the position of the coincidence $\boldsymbol{x}_{\rm max}$.
        \item A combination of p-values using Fisher's method \citep{fisher}. The p-values can be, for example, how likely the individual event is from background, or how likely is the overlap between events just random. 
    \end{itemize}
    \item Calculate the false-alarm rate of the coincidence as a function of the ranking statistic.
\end{enumerate}

The false-alarm rate (FAR) is built by simulating random coincidences and generating a distribution of the ranking statistic. 
The analysis \textit{Fermi}-ANTARES \citep{antares_fermi} and HAWC-IceCube \citep{hawc_icecube} are, at the time of writing, running in real-time.
The latency of these analyses are in the order of hours: 1 to 12 hours when data is downlinked from the \textit{Fermi}-LAT instrument, and 3 to 7 hours when the data is collected by the HAWC Observatory (which corresponds to the time a point in the sky transits the detector's field of view). 
The calculation of the ranking statistic and the alert sending time is less than a minute. For the public real-time system we have set, with an agreement between the observatories, a threshold of FAR $<4$ per year to send alerts. Alerts are sent as notices and circulars through GCN\footnote{The NuEM notices are found in \url{https://gcn.gsfc.nasa.gov/gcn/amon_nu_em_coinc_events.html}}.

\section{Results}
Table \ref{tab:nuem_alerts} shows the public alerts that appear in GCN, as well as the results of the archival coincidence searches performed so far. 
For the archival coincidences we show the ones that have a FAR $<1$ per year. 

\begin{table}[!htb]
    \centering
    \setlength{\arrayrulewidth}{0.5mm}
    \begin{tabular}{|c|c|c|c|c|c|}
        \hline
        \textbf{Name} &  \textbf{R.A. [$^{\circ}$]} & \textbf{Decl. [$^{\circ}$]} & \textbf{$\delta \theta$ [$^{\circ}$]} & \textbf{FAR [$yr^{-1}$]} & \textbf{Time UTC} \\
        \hline
        \rowcolor{lightgray} \multicolumn{6}{|c|}{Real-time alerts} \\
        \hline
        NuEM-210515A & 93.64 & 14.66 & 0.15 & 3.93 & 2021-05-15 00:20:43 \\
        NuEM-210515B & 93.93 & 12.51 & 0.20 & 1.90 & 2021-05-15 00:19:27 \\
        NuEM-210111A & 162.34 & 19.46 & 0.37 & 3.85 & 2021-01-11 13:06:41 \\
        NuEM-201124A & 134.99 & 7.74 & 0.23 & 2.96 & 2020-11-24 14:13:37 \\
        NuEM-201107A & 140.20 & 29.76 & 0.15 & 3.49 & 2020-11-07 15:55:31 \\
        ANTARES-Fermi 200704A & 255.42 & -34.48 & 0.43 & 0.98 & 2020-07-04 15:53:48 \\
        NuEM-200202A & 200.30 & 12.71 & 0.17 & 1.39 & 2020-02-02 14:07:52 \\
        ANTARES-Fermi 191011A & 49.96 & 18.80 & 0.40 & 1.21 & 2019-10-11 15:54:32 \\
        \hline
        \rowcolor{lightgray} \multicolumn{6}{|c|}{Archival Coincidences} \\
        \hline
        ANTARES-Fermi & 248.00 & -7.7 & 0.07 & 0.09 & 2012-11-21  20:19:52 \\
        ANTARES-Fermi & 279.68 & -5.05 & 0.10 & 0.09 & 2014-08-05 11:13:33 \\
        HAWC-IceCube & 4.93 & 2.96 & 0.16 & 0.99 & 2016-12-12 04:38:41 \\
        HAWC-IceCube & 173.99 & 2.27 & 0.53 & 0.026 & 2018-04-12 07:54:51 \\
        HAWC-ANTARES & 25.6 & 25.0 & 0.2 & 0.7 & 2016-01-08 04:39:38 \\
        HAWC-ANTARES & 222.8 & -0.8 & 0.2 & 0.87 & 2017-09-07 01:21:22 \\
        HAWC-ANTARES & 85.4 & 3.4 & 0.2 & 0.41 & 2019-03-29 03:01:18 \\
        \hline
    \end{tabular}
    \caption{Real-time and Archival coincidences in the AMON NuEM channel. Real-time alerts are sent when the FAR is less than 4 per year. Archival coincidences have a FAR of less than 1 per year. Although the ANTARES-Fermi alerts are part of the AMON NuEM channel, the alerts are currently sent via GCN circulars instead of notices.}
    \label{tab:nuem_alerts}
\end{table}

Follow-up observations of the real-time alerts have been made by a couple of observatories. Table \ref{tab:followups} shows some of the instruments that submitted GCN circulars of their follow-up observations. 

We have also looked at the Fermi All-sky Variability Analysis (FAVA)\footnote{\url{https://fermi.gsfc.nasa.gov/ssc/data/access/lat/FAVA/}} as well as the SIMBAD\footnote{\url{http://simbad.u-strasbg.fr/simbad/}} and NED\footnote{\url{http://ned.ipac.caltech.edu/}} catalogs, but no counterpart has been found so far for any of the coincidence alerts. 

\begin{table}[!htb]
    \centering
    \setlength{\arrayrulewidth}{0.5mm}
    \begin{tabular}{|c|c|}
        \hline
        \textbf{Name} & \textbf{Followed by}\\
        \hline
        \href{https://gcn.gsfc.nasa.gov/gcn3/30010.gcn3}{NuEM-210515A/B} & \href{https://gcn.gsfc.nasa.gov/gcn3/30024.gcn3}{ANTARES} \\ \href{https://gcn.gsfc.nasa.gov/gcn3/30010.gcn3}{NuEM-210111A} & \href{https://gcn.gsfc.nasa.gov/gcn3/29294.gcn3}{ANTARES}, \href{https://gcn.gsfc.nasa.gov/gcn3/29288.gcn3}{INTEGRAL},\href{https://gcn.gsfc.nasa.gov/gcn3/29298.gcn3}{MAXI}\\
        \href{https://gcn.gsfc.nasa.gov/gcn3/28950.gcn3}{NuEM-201124A} & \href{https://gcn.gsfc.nasa.gov/gcn3/28953.gcn3}{ANTARES} \\
        \href{https://gcn.gsfc.nasa.gov/gcn3/28865.gcn3}{NuEM-201107A} & \href{https://gcn.gsfc.nasa.gov/gcn3/28884.gcn3}{\textit{Fermi}-LAT} \\
        \href{https://gcn.gsfc.nasa.gov/gcn3/26963.gcn3}{NuEM-200202A} & \href{https://gcn.gsfc.nasa.gov/gcn3/26973.gcn3}{MASTER}, \href{https://gcn.gsfc.nasa.gov/gcn3/26976.gcn3}{ANTARES} \\
        \href{https://gcn.gsfc.nasa.gov/gcn3/26005.gcn3}{FERMI-ANTARES-191011A} & \href{https://gcn.gsfc.nasa.gov/gcn3/26006.gcn3}{MASTER} \\
        \hline
    \end{tabular}
    \caption{Follow-up observations of the NuEM alerts published as GCN Circulars. No counterpart was found in these observations. If seen online, each alert has a link to their respective circular.}
    \label{tab:followups}
\end{table}

\section{Summary}

The multi-messenger approach has become a new paradigm to study astrophysical objects. Its advantages are that we can have a better understanding of different phenomena in the universe, such as sources of high-energy cosmic rays and neutrinos, the processes leading to gamma-ray bursts and the nature of dark matter. However, the tasks are challenging and require development of new techniques as well as the cooperation between different collaborations. AMON is a framework that aims to help with these issues. As an example, we have shown here the NuEM channel, a channel that performs coincidence analyses between neutrino data and gamma-ray data. The datasets used are diverse since they come from observatories such as IceCube, ANTARES, HAWC and Fermi-LAT. The AMON-NuEM channel has been sending alerts to the astrophysical community since 2019.  The channel is still growing, with new coincidences planned to be added to it such as the HAWC-ANTARES analysis.  

\bibliographystyle{JHEP}
\bibliography{biblio}


\clearpage
\input{antares_authors}

\input{hawc_authors}
\input{icecube_authors}
%
%
%

\end{document}

%% file: antares_authors.tex
\section*{Full Authors List: ANTARES Collaboration}

\scriptsize
\noindent
A.~Albert$^{1,2}$,
S.~Alves$^{3}$,
M.~Andr\'e$^{4}$,
M.~Anghinolfi$^{5}$,
G.~Anton$^{6}$,
M.~Ardid$^{7}$,
S.~Ardid$^{7}$,
J.-J.~Aubert$^{8}$,
J.~Aublin$^{9}$,
B.~Baret$^{9}$
% ...
\\

\noindent
$^1$Universit\'e de Strasbourg, CNRS,  IPHC UMR 7178, F-67000 Strasbourg, France.
$^2$ Universit\'e de Haute Alsace, F-68100 Mulhouse, France.
$^3$IFIC - Instituto de F\'isica Corpuscular (CSIC - Universitat de Val\`encia) c/ Catedr\'atico Jos\'e Beltr\'an, 2 E-46980 Paterna, Valencia, Spain.
$^4$Technical University of Catalonia, Laboratory of Applied Bioacoustics, Rambla Exposici\'o, 08800 Vilanova i la Geltr\'u, Barcelona, Spain.
$^5$INFN - Sezione di Genova, Via Dodecaneso 33, 16146 Genova, Italy.
$^6$Friedrich-Alexander-Universit\"at Erlangen-N\"urnberg, Erlangen Centre for Astroparticle Physics, Erwin-Rommel-Str. 1, 91058 Erlangen, Germany.
$^7$Institut d'Investigaci\'o per a la Gesti\'o Integrada de les Zones Costaneres (IGIC) - Universitat Polit\`ecnica de Val\`encia. C/  Paranimf 1, 46730 Gandia, Spain.
$^8$Aix Marseille Univ, CNRS/IN2P3, CPPM, Marseille, France.
$^9$Universit\'e de Paris, CNRS, Astroparticule et Cosmologie, F-75013 Paris, France.
% ...

%% file: hawc_authors.tex
\section*{Full Authors List: HAWC Collaboration}

\scriptsize
\noindent
A.U. Abeysekara$^{48}$,
A. Albert$^{21}$,
R. Alfaro$^{14}$,
C. Alvarez$^{41}$,
J.D. Álvarez$^{40}$,
J.R. Angeles Camacho$^{14}$,
J.C. Arteaga-Velázquez$^{40}$,
K. P. Arunbabu$^{17}$,
D. Avila Rojas$^{14}$,
H.A. Ayala Solares$^{28}$
% ...

\noindent
% ...
$^{14}$Instituto de Física, Universidad Nacional Autónoma de México, Ciudad de México, México,
% ...
$^{17}$Instituto de Geofísica, Universidad Nacional Autónoma de México, Ciudad de México, México,
% ...
$^{21}$Physics Division, Los Alamos National Laboratory, Los Alamos, NM, USA,
% ...
$^{28}$Department of Physics, Pennsylvania State University, University Park, PA, USA,
% ...
$^{40}$Instituto de Física y Matemáticas, Universidad Michoacana de San Nicolás de Hidalgo, Morelia, Michoacán, México,
$^{41}$FCFM-MCTP, Universidad Autónoma de Chiapas, Tuxtla Gutiérrez, Chiapas, México,
% ...
$^{48}$Department of Physics and Astronomy, University of Utah, Salt Lake City, UT, USA,
% ...

\subsection*{HAWC Acknowledgments}
\noindent
We acknowledge the support from: the US National Science Foundation (NSF); the US Department of Energy Office of High-Energy Physics; the Laboratory Directed Research and Development (LDRD) program of Los Alamos National Laboratory; Consejo Nacional de Ciencia y Tecnolog\'ia (CONACyT), M\'exico, grants 271051, 232656, 260378, 179588, 254964, 258865, 243290, 132197, A1-S-46288, A1-S-22784, c\'atedras 873, 1563, 341, 323, Red HAWC, M\'exico; DGAPA-UNAM grants IG101320, IN111716-3, IN111419, IA102019, IN110621, IN110521; VIEP-BUAP; PIFI 2012, 2013, PROFOCIE 2014, 2015; the University of Wisconsin Alumni Research Foundation; the Institute of Geophysics, Planetary Physics, and Signatures at Los Alamos National Laboratory; Polish Science Centre grant, DEC-2017/27/B/ST9/02272; Coordinaci\'on de la Investigaci\'on Cient\'ifica de la Universidad Michoacana; Royal Society - Newton Advanced Fellowship 180385; Generalitat Valenciana, grant CIDEGENT/2018/034; Chulalongkorn University's CUniverse (CUAASC) grant; Coordinaci\'on General Acad\'emica e Innovaci\'on (CGAI-UdeG), PRODEP-SEP UDG-CA-499; Institute of Cosmic Ray Research (ICRR), University of Tokyo, H.F. acknowledges support by NASA under award number 80GSFC21M0002. We also acknowledge the significant contributions over many years of Stefan Westerhoff, Gaurang Yodh and Arnulfo Zepeda Dominguez, all deceased members of the HAWC collaboration. Thanks to Scott Delay, Luciano D\'iaz and Eduardo Murrieta for technical support.

%% file: icecube_authors.tex
\section*{Full Authors List: IceCube Collaboration}

\scriptsize
\noindent
R. Abbasi$^{17}$,
M. Ackermann$^{59}$,
J. Adams$^{18}$,
J. A. Aguilar$^{12}$,
M. Ahlers$^{22}$,
M. Ahrens$^{50}$,
C. Alispach$^{28}$,
A. A. Alves Jr.$^{31}$,
N. M. Amin$^{42}$,
R. An$^{14}$
% ...
\\

\noindent
% ...
$^{12}$ Universit{\'e} Libre de Bruxelles, Science Faculty CP230, B-1050 Brussels, Belgium \\
% ...
$^{14}$ Department of Physics and Laboratory for Particle Physics and Cosmology, Harvard University, Cambridge, MA 02138, USA \\
% ...
$^{17}$ Department of Physics, Loyola University Chicago, Chicago, IL 60660, USA \\
$^{18}$ Dept. of Physics and Astronomy, University of Canterbury, Private Bag 4800, Christchurch, New Zealand \\
% ...
$^{22}$ Niels Bohr Institute, University of Copenhagen, DK-2100 Copenhagen, Denmark \\
% ...
$^{28}$ D{\'e}partement de physique nucl{\'e}aire et corpusculaire, Universit{\'e} de Gen{\`e}ve, CH-1211 Gen{\`e}ve, Switzerland \\
% ...
$^{31}$ Karlsruhe Institute of Technology, Institute for Astroparticle Physics, D-76021 Karlsruhe, Germany  \\
% ...
$^{42}$ Bartol Research Institute and Dept. of Physics and Astronomy, University of Delaware, Newark, DE 19716, USA \\
% ...
$^{50}$ Oskar Klein Centre and Dept. of Physics, Stockholm University, SE-10691 Stockholm, Sweden \\
% ...
$^{59}$ DESY, D-15738 Zeuthen, Germany \\
% ...

\subsection*{IceCube Acknowledgments}
\noindent
USA {\textendash} U.S. National Science Foundation-Office of Polar Programs,
U.S. National Science Foundation-Physics Division,
U.S. National Science Foundation-EPSCoR,
Wisconsin Alumni Research Foundation,
Center for High Throughput Computing (CHTC) at the University of Wisconsin{\textendash}Madison,
Open Science Grid (OSG),
Extreme Science and Engineering Discovery Environment (XSEDE),
Frontera computing project at the Texas Advanced Computing Center,
U.S. Department of Energy-National Energy Research Scientific Computing Center,
Particle astrophysics research computing center at the University of Maryland,
Institute for Cyber-Enabled Research at Michigan State University,
and Astroparticle physics computational facility at Marquette University;
Belgium {\textendash} Funds for Scientific Research (FRS-FNRS and FWO),
FWO Odysseus and Big Science programmes,
and Belgian Federal Science Policy Office (Belspo);
Germany {\textendash} Bundesministerium f{\"u}r Bildung und Forschung (BMBF),
Deutsche Forschungsgemeinschaft (DFG),
Helmholtz Alliance for Astroparticle Physics (HAP),
Initiative and Networking Fund of the Helmholtz Association,
Deutsches Elektronen Synchrotron (DESY),
and High Performance Computing cluster of the RWTH Aachen;
Sweden {\textendash} Swedish Research Council,
Swedish Polar Research Secretariat,
Swedish National Infrastructure for Computing (SNIC),
and Knut and Alice Wallenberg Foundation;
Australia {\textendash} Australian Research Council;
Canada {\textendash} Natural Sciences and Engineering Research Council of Canada,
Calcul Qu{\'e}bec, Compute Ontario, Canada Foundation for Innovation, WestGrid, and Compute Canada;
Denmark {\textendash} Villum Fonden and Carlsberg Foundation;
New Zealand {\textendash} Marsden Fund;
Japan {\textendash} Japan Society for Promotion of Science (JSPS)
and Institute for Global Prominent Research (IGPR) of Chiba University;
Korea {\textendash} National Research Foundation of Korea (NRF);
Switzerland {\textendash} Swiss National Science Foundation (SNSF);
United Kingdom {\textendash} Department of Physics, University of Oxford.

%% file: main.bbl
\providecommand{\href}[2]{#2}\begingroup\raggedright\begin{thebibliography}{10}

\bibitem{bns}
B.P.~{Abbott}, R.~{Abbott}, T.D.~{Abbott} et~al., \emph{{Multi-messenger
  Observations of a Binary Neutron Star Merger}},
  \href{https://doi.org/10.3847/2041-8213/aa91c9}{\emph{ApJ} {\bfseries 848}
  (2017) L12}.

\bibitem{icneutrino}
M.G.~{Aartsen}, M.~{Ackermann}, J.~{Adams} et~al., \emph{{Multimessenger
  observations of a flaring blazar coincident with high-energy neutrino
  IceCube-170922A}},
  \href{https://doi.org/10.1126/science.aat1378}{\emph{Science} {\bfseries 361}
  (2018) 8}.

\bibitem{mmareview}
F.~Halzen, \emph{{Multi-messenger astronomy: cosmic rays, gamma-rays, and
  neutrinos}},  in \emph{{21st Texas Symposium on Relativistic Astrophysics
  (Texas in Tuscany)}}, 2, 2003,
  \href{https://doi.org/10.1142/9789812704009_0011}{DOI}.

\bibitem{mmareview2}
A.~Neronov, \emph{{Introduction to multi-messenger astronomy}},
  \href{https://doi.org/10.1088/1742-6596/1263/1/012001}{\emph{J. Phys. Conf.
  Ser.} {\bfseries 1263} (2019) 012001}
  [\href{https://arxiv.org/abs/1907.07392}{{\ttfamily 1907.07392}}].

\bibitem{amon_2020}
H.A.~{Ayala Solares}, S.~Coutu, D.F.~Cowen et~al., \emph{The {Astrophysical}
  {Multimessenger} {Observatory} {Network} ({AMON}): {Performance} and science
  program},
  \href{https://doi.org/https://doi.org/10.1016/j.astropartphys.2019.06.007}{\emph{Astroparticle
  Physics} {\bfseries 114} (2020) 68}.

\bibitem{neutrinoProd}
D.~{Biehl}, D.~{Boncioli}, A.~{Fedynitch} et~al., \emph{{Astrophysical neutrino
  production and impact of associated uncertainties in photo-hadronic
  interactions of UHECRs}},  in \emph{European Physical Journal Web of
  Conferences}, vol.~208 of \emph{European Physical Journal Web of
  Conferences}, p.~04002, May, 2019,
  \href{https://doi.org/10.1051/epjconf/201920804002}{DOI}.

\bibitem{icecube}
M.G.~{Aartsen}, M.~{Ackermann}, J.~{Adams} et~al., \emph{{The IceCube Neutrino
  Observatory: Instrumentation and Online Systems}},
  \href{https://doi.org/10.1088/1748-0221/12/03/P03012}{\emph{\emph{JINST}}
  {\bfseries P03012} (2017) 12}.

\bibitem{antares}
M.~Ageron, J.~Aguilar, I.A.~Samarai et~al., \emph{{ANTARES: The first undersea
  neutrino telescope}}, {\emph{Nucl. Instrum. Methods Phys. Res. A} {\bfseries
  656} (2011) 11}.

\bibitem{hawc}
A.U.~{Abeysekara}, A.~{Albert}, R.~{Alfaro} et~al., \emph{{Observation of the
  Crab Nebula with the HAWC Gamma-Ray Observatory}}, {\emph{ApJ} {\bfseries
  843} (2017) 39}.

\bibitem{fermi}
M.~Ackermann, M.~Ajello, A.~Albert et~al., \emph{{The \textit{Fermi} Large Area
  Telescope on Orbit: Event Classification,Instrument Response Functions, and
  Calibration}}, {\emph{ApJS} {\bfseries 203} (2012) 4}.

\bibitem{antares_fermi}
H.A.~{Ayala Solares}, S.~{Coutu}, J.J.~{DeLaunay} et~al., \emph{{A Search for
  Cosmic Neutrino and Gamma-Ray Emitting Transients in 7.3 Years of ANTARES and
  Fermi LAT Data}},
  \href{https://doi.org/10.3847/1538-4357/ab4a74}{\emph{Astrophys. J.}
  {\bfseries 886} (2019) 98}
  [\href{https://arxiv.org/abs/1904.06420}{{\ttfamily 1904.06420}}].

\bibitem{hawc_icecube}
H.A.~{Ayala Solares}, S.~{Coutu}, J.J.~{DeLaunay} et~al., \emph{{Multimessenger
  Gamma-Ray and Neutrino Coincidence Alerts Using HAWC and IceCube Subthreshold
  Data}}, \href{https://doi.org/10.3847/1538-4357/abcaa4}{\emph{ApJ} {\bfseries
  906} (2021) 63} [\href{https://arxiv.org/abs/2008.10616}{{\ttfamily
  2008.10616}}].

\bibitem{icecube_fermi}
C.F.~Turley, D.B.~Fox, A.~Keivani, J.J.~DeLaunay, D.F.~Cowen, M.~Mostafa
  et~al., \emph{{A Coincidence Search for Cosmic Neutrino and Gamma-Ray
  Emitting Sources Using IceCube and $Fermi$ LAT Public Data}},
  \href{https://doi.org/10.3847/1538-4357/aad195}{\emph{Astrophys. J.}
  {\bfseries 863} (2018) 64}
  [\href{https://arxiv.org/abs/1802.08165}{{\ttfamily 1802.08165}}].

\bibitem{fisher}
R.A.~Fisher, \emph{Statistical methods for research workers}, Edinburgh, Oliver
  and Boyd (1938).

\end{thebibliography}\endgroup
